\documentstyle[12pt]{article}

\begin{document}


\def\a{\alpha}
\def\b{\beta}
\def\c{\varepsilon}
\def\d{\delta}
\def\e{\epsilon}
\def\f{\phi}
\def\g{\gamma}
\def\h{\theta}
\def\k{\kappa}
\def\l{\lambda}
\def\m{\mu}
\def\n{\nu}
\def\p{\psi}
\def\q{\partial}
\def\r{\rho}
\def\s{\sigma}
\def\t{\tau}
\def\u{\upsilon}
\def\v{\varphi}
\def\w{\omega}
\def\x{\xi}
\def\y{\eta}
\def\z{\zeta}
\def\D{\Delta}
\def\G{\Gamma}
\def\L{\Lambda}
\def\F{\Phi}
\def\P{\Psi}
\def\S{\Sigma}

\def\o{\over}
\def\beq{\begin{eqnarray}}
\def\eeq{\end{eqnarray}}
\newcommand{\gsim}{ \mathop{}_{\textstyle \sim}^{\textstyle >} }
\newcommand{\lsim}{ \mathop{}_{\textstyle \sim}^{\textstyle <} }

\def\IJMP{Int.~J.~Mod.~Phys. }
\def\MPL{Mod.~Phys.~Lett. }
\def\NP{Nucl.~Phys. }
\def\PL{Phys.~Lett. }
\def\PR{Phys.~Rev. }
\def\PRL{Phys.~Rev.~Lett. }
\def\PTP{Prog.~Theor.~Phys. }
\def\ZP{Z.~Phys. }


\baselineskip 0.7cm

\begin{titlepage}
\begin{flushright}
UT-894
\\
May, 2000
\end{flushright}

\vskip 1.35cm
\begin{center}
{\large \bf
Quintessential Vanishing of the Cosmological Constant
}
\vskip 1.2cm
Izawa K.-I.
\vskip 0.4cm

{\it Department of Physics and RESCEU, University of Tokyo,\\
     Tokyo 113-0033, Japan}

\vskip 1.5cm

\abstract{
We consider an effective field theory description of gravity
coupled to a scalar field
with volume-preserving diffeomorphism and Weyl invariances.
The smallness of the cosmological constant is achieved when
the potential of the scalar is extremely flat.
This implies a possible relationship between the smallness
and the quintessential sector of the cosmological constant.
}
\end{center}
\end{titlepage}

\setcounter{page}{2}


The smallness of the cosmological constant
poses a severe problem
\cite{Wei,Car}
on our natural understanding
of an effective field theory description of the universe.
The problem is twofold:
One is the apparent absence or cancellation of the contributions
from the standard model dynamics including gravity
to the vacuum energy. The other is the small (vanishing or
tiny nonvanishing) value of the cosmological constant itself.

The former implies that we need another 
compensating sector whose contribution
to the vacuum energy is a dominant one and controls the size of the
net cosmological constant.
The simplest candidate description of the additional sector is given
by the interaction of a scalar field in four dimensions.
If the dynamics of this scalar field by itself adjusted
the value of the cosmological constant to (almost) zero,
it would also resolve the latter half of the problem.
Unfortunately, this turns out to be not the case generally
in the ordinary Einstein gravity
\cite{Wei,Dol}.

If it is nonvanishing as recent observations might suggest
\cite{Car},
the size of the cosmological constant may be determined by
a quintessential scalar field whose potential is extremely flat
\cite{Car,Bin}.
In this paper, we investigate a possible relationship
between the smallness and the quintessential sector of the
cosmological constant.
Concretely, we consider a scalar field in an alternative Einstein gravity
\cite{Iza}
with the cosmological constant as an integration constant
determined by an initial condition of the universe.
This setup
eliminates the former half of the problem.
As for the latter half, it
sheds new light on a possible
solution to the problem
and implies that the scalar may be quintessential,
though we do not completely avoid fine tuning of the scalar interaction
to achieve a tiny value of the cosmological constant in accordance
with quintessential difficulty
\cite{Kol}.

Let us introduce a symmetric tensor $g_{\mu \nu}$
and a scalar $\f$
in four-dimensional spacetime.
We impose volume-preserving diffeomorphism and Weyl invariances
to obtain the Lagrangian
\cite{Iza}.
Weyl symmetry
\beq
  \d_{_W} g_{\m \n} = \l g_{\m \n}, \quad \d_{_W} \f = 0
\eeq
implies that the metric field comes in the Lagrangian
solely in the combination
\beq
  {\bar g}_{\m \n} = g^{-{1 \over 4}} g_{\m \n},
\eeq
where $g = -\det g_{\m \n}$.
Invariance under volume-preserving diffeomorphism
\beq
  \d_{_V} {\bar g}_{\m \n} = -\c^\r \q_\r {\bar g}_{\m \n}
                      - {\bar g}_{\m \r} \q_\n \c^\r
                      - {\bar g}_{\n \r} \q_\m \c^\r, \quad
  \d_{_V} \f = -\c^\m \q_\m \f; \quad
  \q_\m \c^\m = 0
\eeq
can be achieved
\cite{Buc}
in complete analogy to
the case of full diffeomorphism invariance.
The Lagrangian bears the same form as
that in the ordinary Einstein gravity except for the field $g_{\m \n}$
in the latter replaced by ${\bar g}_{\m \n}$ in the former:
\beq
  {\cal L} = {1 \o 2}{\bar g}^{\m \n}\q_\m \f \q_\n \f - V(\f)
   + {1 \o 2}U(\f){\bar R}.
\eeq
Here ${\bar g}^{\m \n}$ denotes the inverse
of ${\bar g}_{\m \n}$,
$\bar R$ is the scalar curvature corresponding to ${\bar g}_{\m \n}$,
and we adopt the mass unit based on a fundamental scale
(presumably around the Planck scale) in some fundamental theory,
which yields the present effective field theory of gravity
\footnote{Quantization as an effective field theory is
straightforward in the present formulation
\cite{Iza}.}
in a portion of its moduli space parameterizing effective theories.
The volume density
(in particular, the apparent cosmological term)
is absent from the expression
since $\det {\bar g}_{\m \n} = -1$.
Note that the functions $V(\f)$ and $U(\f)$
have independent meanings in the above Lagrangian due to the Weyl symmetry.
This is in sharp contrast to the case of the ordinary
Einstein gravity, where the function corresponding to $U(\f)$
may be brought to one by a metric redefinition through
a $\f$-dependent Weyl transformation.

Let $S[{\bar g}_{\m \n}, \f]$ be the corresponding action,
whose equations of motion
are obtained as
\beq
  {\d S \over \d g_{\m \n}}
   = g^{-{1 \over 4}}({\d S \over \d {\bar g}_{\m \n}}
    -{1 \over 4}{\bar g}^{\m \n}{\bar g}_{\r \s}
     {\d S \over \d {\bar g}_{\r \s}}) = 0, \quad
  {\d S \over \d \f} = 0.
\eeq
On the other hand, we get an equation
\beq
  {\bar D}_\m {\d S \over \d {\bar g}_{\m \n}} = 0
\eeq
with the aid of the second equation of motion and
the Noether identity due to diffeomorphism invariance
of the action $S[g_{\m \n}, \f]$,
where ${\bar D}_\m$ denotes the covariant derivative
corresponding to ${\bar g}_{\m \n}$.
Hence the first equation of motion yields
\beq
  {\bar D}_\m ({\d S \over \d {\bar g}_{\m \n}}
   -{1 \over 4}{\bar g}^{\m \n}{\bar g}_{\r \s}
    {\d S \over \d {\bar g}_{\r \s}})
   = -{1 \over 4}{\bar g}^{\m \n} \q_\m ({\bar g}_{\r \s}
    {\d S \over \d {\bar g}_{\r \s}}) = 0,
\eeq
which indicates that
\beq
  {\bar g}_{\r \s}{\d S \over \d {\bar g}_{\r \s}} = 2\L
\eeq
is a constant independent of spacetime.
Therefore we conclude that the equations of motion are given by
\beq
  {\d S \over \d {\bar g}_{\m \n}} -{1 \o 2}\L {\bar g}^{\m \n} = 0, \quad
  {\d S \over \d \f} = 0,
\eeq
which are none other than those in the ordinary Einstein gravity
with a partial gauge-fixing $g = 1$.
This implies that the theory is equivalent
to the ordinary Einstein gravity with the cosmological constant
as an integration constant determined by an initial condition.%
\footnote{We may use the Weyl transformation to impose
the unimodular condition $g = 1$ in the theory.
Then the theory is classically reduced to the unimodular gravity
\cite{Wei,Bij},
which is known to have the features just mentioned.}

In particular, the background with a constant scalar field (after inflation)
satisfies
\beq
  U(\f){\bar R} = V(\f)+\L, \quad
  {1 \o 2}{\q U(\f) \o \q \f}{\bar R} = {\q V(\f) \o \q \f}.
\eeq
This indicates that an almost flat background metric is realized
provided $U'(\f) \neq 0$ and $V'(\f)/U'(\f)$ is small enough,
which implies that the potential $V(\f)$ is quintessential.
For a small $V(\f)+\L$, we may have a stable vacuum with a small cosmological
constant at a reasonable value of the field $\f$.
If $V(\f)+\L$ were not small so that $U(\f)$ should be
too large to attain small ${\bar R}$, the
effective field theory description would break down.%
\footnote{In the ordinary Einstein gravity, the vacuum energy
is not expected to be small without fine tuning of couplings.
Then $U(\f)$,
if present, is enforced to become large, which implies decoupling
of gravity
\cite{Wei}.}
Hence the present setup implies a tiny cosmological
constant in its range of validity.

The vacuum energy $V(\f)+\L$ is determined by the initial condition
of the universe. The quantum state of the universe may be a
superposition of states with different values of $\L$, and
a `chaotic initial condition
\cite{Lin}'
for $\L$ is expected through decoherence.
Then the classical patches in the universe realize
various values of $\L$, which contain a desirable small value of
$V(\f)+\L$ described by the present effective field
theory of gravity. On the other hand, a portion of the moduli space
corresponding to large values of $V(\f)+\L$ in the
fundamental theory
would deserve other descriptions generically with other field contents,
other spacetime dimensions and so forth.
Thus the presented setup has chances to achieve an almost flat
four-dimensional spacetime.

In this paper, we have utilized an alternative Einstein gravity
with a cosmological constant as an integration constant.
A similar setup may be realized in a higher-dimensional
gravity with warped compactification
\cite{Rub}.
Combined with a brane world scenario,%
\footnote{For investigations on the vanishing cosmological constant
in brane world scenarios, see
Ref.\cite{Ark}.}
the four-dimensional metric
induced on a brane from a bulk spacetime might have meanings
independent of the field-dependent Weyl rescalings on the brane.
Then the quintessential vanishing of the effective four-dimensional
cosmological constant may work just as in the alternative
Einstein gravity with a quintessential scalar field.

\section*{Acknowledgments}

The author would like to thank Y.~Nomura, T.~Watari and T.~Yanagida
for valuable discussions.

\end{document}